%% file: aries1.tex
\input balticw
\input psfig.sty
\def\ts{\thinspace}

\year 2002
\VOLUME{~11}
\PAGES{219--229}
\pageno=219

\TITLE={PHOTOMETRIC INVESTIGATION OF THE \hfil\break MBM 12 MOLECULAR
CLOUD AREA IN ARIES.\hfil\break  I. PHOTOELECTRIC PHOTOMETRY}

\AUTHOR= {A.  Kazlauskas, K. \v{C}ernis, V. Laugalys and V.
Strai\v{z}ys}

\ARTHEAD={MBM 12 molecular cloud area in Aries. I. Photometry}

\AUTHORHEAD={A. Kazlauskas, K. \v{C}ernis, V. Laugalys, V. Strai\v{z}ys}

\ADDRESS={Institute of Theoretical Physics and Astronomy,
Go\v{s}tauto 12,\hfil\break Vilnius 2600, Lithuania;
e-mail addresses: algisk@itpa.lt, cernis@itpa.lt, vygandas@itpa.lt,
straizys@itpa.lt}

\SUBMITTED={May 25, 2002}

\SUMMARY={The results of photoelectric photometry in the {\itl Vilnius}
seven-color system are given for 152 stars down to 12.2 mag in the area
of the molecular cloud MBM 12 and the dust clouds L1454 and L1457 in
Aries.  The results of photometric classification of stars are also
given.  The investigation of interstellar extinction in the area is
described in the next paper.}

\KEYWORDS={techniques:  photometric:  Vilnius photometric system --
stars:  fundamental parameters -- molecular clouds:  individual:  MBM 12
-- dust clouds:  individual:  L1454 and L1457}

\printheader

\section{1. INTRODUCTION}

This paper extends the program of investigation of the
Taurus--Auriga--Perseus--Aries dark clouds by photoelectric photometry
in the Vilnius seven-color system (Strai\v zys 1977, 1992).
\vskip0.5mm

The area investigated in this paper includes the molecular cloud MBM 12
(Magnani, Blitz \& Mundy 1985) and is bounded by the following 2000.0
coordinates:  RA from $2^{\rm h}51^{\rm m}$ to $3^{\rm h}01^{\rm m}$ and
DEC from +18$\degr$30$\arcmin$ to +21$\degr$30$\arcmin$.  The molecular
cloud coincides with two large dust clouds L1454 and L1457 (Lynds 1962).
Inside the L1454 cloud, two smaller clouds, L1453 and L1458, have been
identified by Lynds.
\vskip0.5mm

In the present paper we describe the methods of observation and
reductions and give the results of photometry and photometric
classification.  The methods of photometric classification of stars and
determination of absolute magnitudes, color excesses, interstellar
extinctions and distances are given in Strai\v zys, \v Cernis,
Kazlauskas, Laugalys (2002, Paper II).  That paper also contains the
investigation of interstellar extinction in the area and the
determination of distance to the dust cloud.

\section{2. OBSERVATIONS, REDUCTIONS AND THE CATALOG}

Photoelectric photometry of 152 stars, identified in Figures 1 and 2, in
the standard passbands of the {\it Vilnius} system was performed in
2000--2001 with the two telescopes:  the 165 cm telescope of the
Mol\.etai Observatory in Lithuania and the 150 cm telescope of the
University of Arizona at Mt. Lemmon, Arizona.  In the area almost all
stars down to 12.2 mag and some fainter stars down to 13.5 mag were
measured.  In all cases we tried to reach the accuracy of flux
measurements better than $\pm$0.01 mag, so at least \hbox{10\ts 000}
pulses were usually counted in each filter.  For the faintest stars this
condition was sometimes disobeyed, and the accuracy of their magnitudes
and color indices can be somewhat lower, especially in the ultraviolet
filters for red or reddened stars.  \vskip0.5mm

The extinction coefficients in the atmosphere for every moment of
observation were determined by the Nikonov (1976) method adjusted to the
{\it Vilnius} system by Zdanavi\v cius (1975, 1996).  In the Mol\.etai
observations, for the extinction determination the stars HD 23609 ($V$ =
6.97, F9\ts IV) in Pleiades and HD 18106 ($V$ = 8.64, G2\ts
V) in the investigated area were used.  In the Arizona observations,
the stars HD 2688 ($V$ = 7.52, F6\ts IV) in SA 44 and HD 27406 ($V$ =
7.48, G0\ts V) in Hyades were used.  Transformation equations from the
instrumental systems to the standard system were determined by
photometry of $\sim$20 stars of various spectral types in the Cygnus
Standard Region (Zdanavi\v cius \& \v Cernien\.e 1985).  \vskip0.5mm

The catalog of the observed stars is given in Table 1. It contains
the star identification number in Figures 1 and 2, the GSC and BD
numbers, right ascensions and declinations, spectral types determined
photometrically, spectral types from other sources (HD, AGK, PPM
catalogs), magnitudes $V$, six color indices and the number of
independent observations for each star, $n$, used to take the average
values of magnitudes and color indices.  The HD and {\it Hipparcos}
(HIP) numbers are given in the Notes.

\WFigure{1}{\psfig{figure=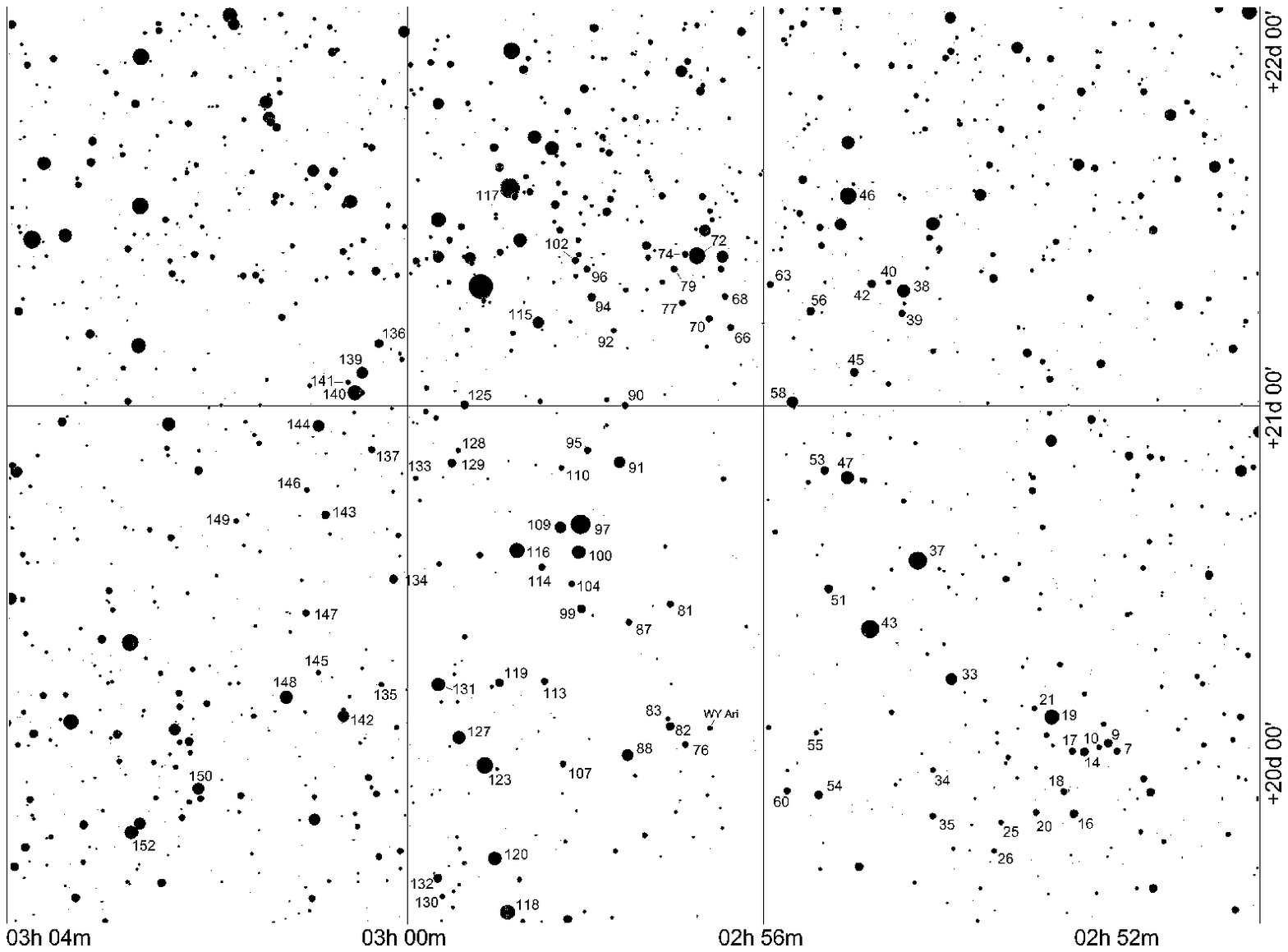,width=12.4truecm,angle=90,clip=}}
{ Identification chart for the northern section of the Aries area. }

\WFigure{2}{\psfig{figure=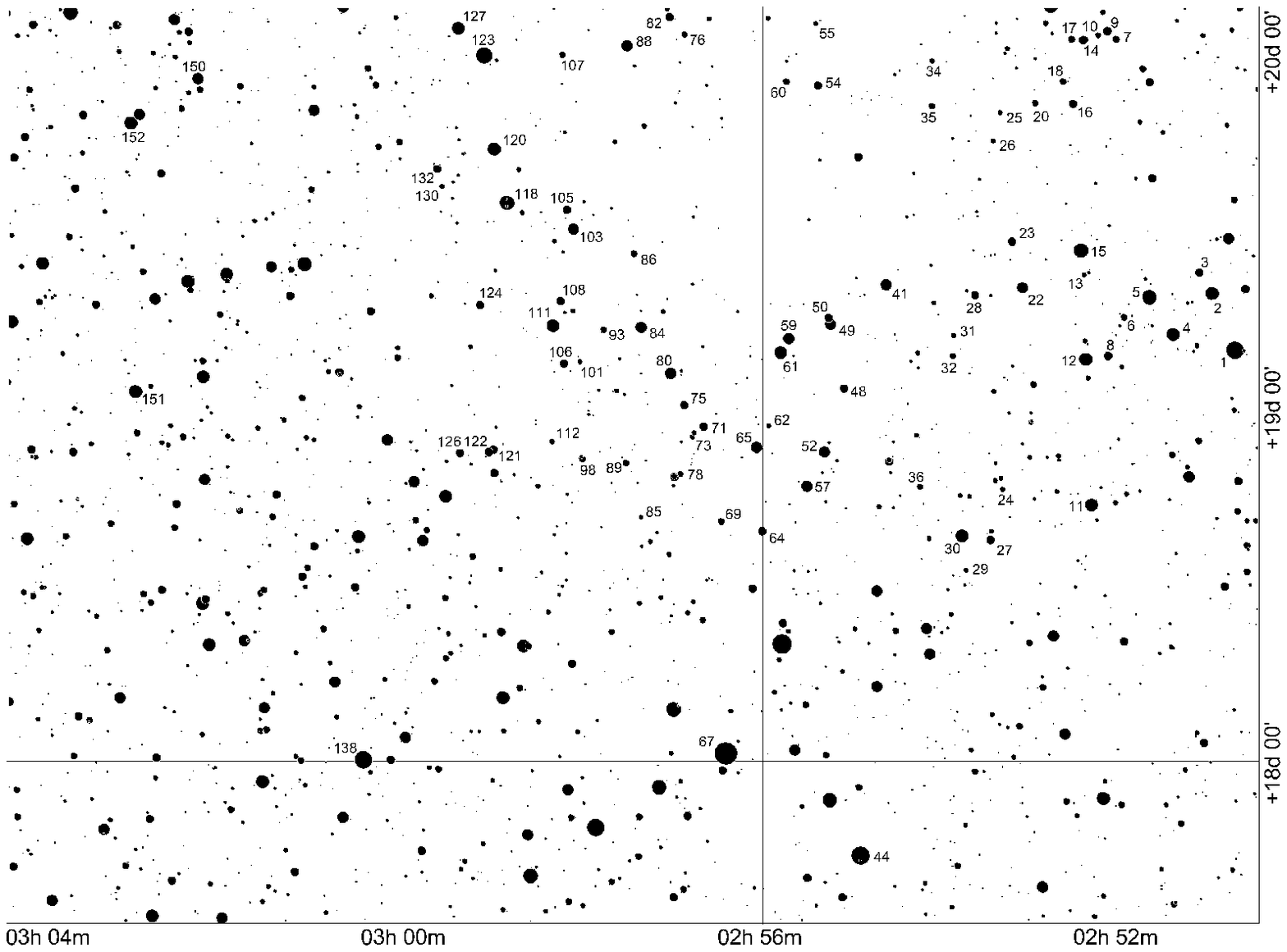,width=12.4truecm,angle=90,clip=}}
{ Identification chart for the southern section of the Aries area. }

\vskip0.5cm

ACKNOWLEDGMENTS.  The investigation is partly supported from the AAS
Chretien Grant of 2000.  We are grateful to the Arizona University
telescope time allocation committee for the observing time on Mount
Lemmon, to the Vatican Observatory community, especially Richard Boyle,
for their help and hospitality during the observing run.  We are also
grateful to K. Zdanavi\v cius and J. Zdanavi\v cius for their help in
the Mol\.etai observations.

\References

\ref Lynds B. T. 1962, ApJS, 7, 1

\ref Magnani L., Blitz L., Mundy L. 1985, ApJ, 295, 402

\ref Nikonov V. B. 1976, Izvestija Crimean Obs., 54, 3

\ref Strai\v zys V. 1977, {\itl Multicolor Stellar Photometry}, Mokslas
Publishers, \hbox {Vilnius,} Lithuania

\ref Strai\v{z}ys V. 1992, {\itl Multicolor Stellar Photometry}, Pachart
Publishing House, Tucson, Arizona

\ref Strai\v zys V., \v Cernis K., Kazlauskas A., Laugalys V. 2002,
Baltic Astro\-nomy, 11, 231 (Paper II, this issue)

\ref Zdanavi\v{c}ius K. 1975, Bull. Vilnius Obs., No. 41, 3

\ref Zdanavi\v{c}ius K. 1996, Baltic Astronomy, 5, 549

\ref Zdanavi\v{c}ius K., \v{C}ernien\.e E. 1985, Bull.  Vilnius Obs.,
No. 69, 3

\vfil\eject

\psfig{figure=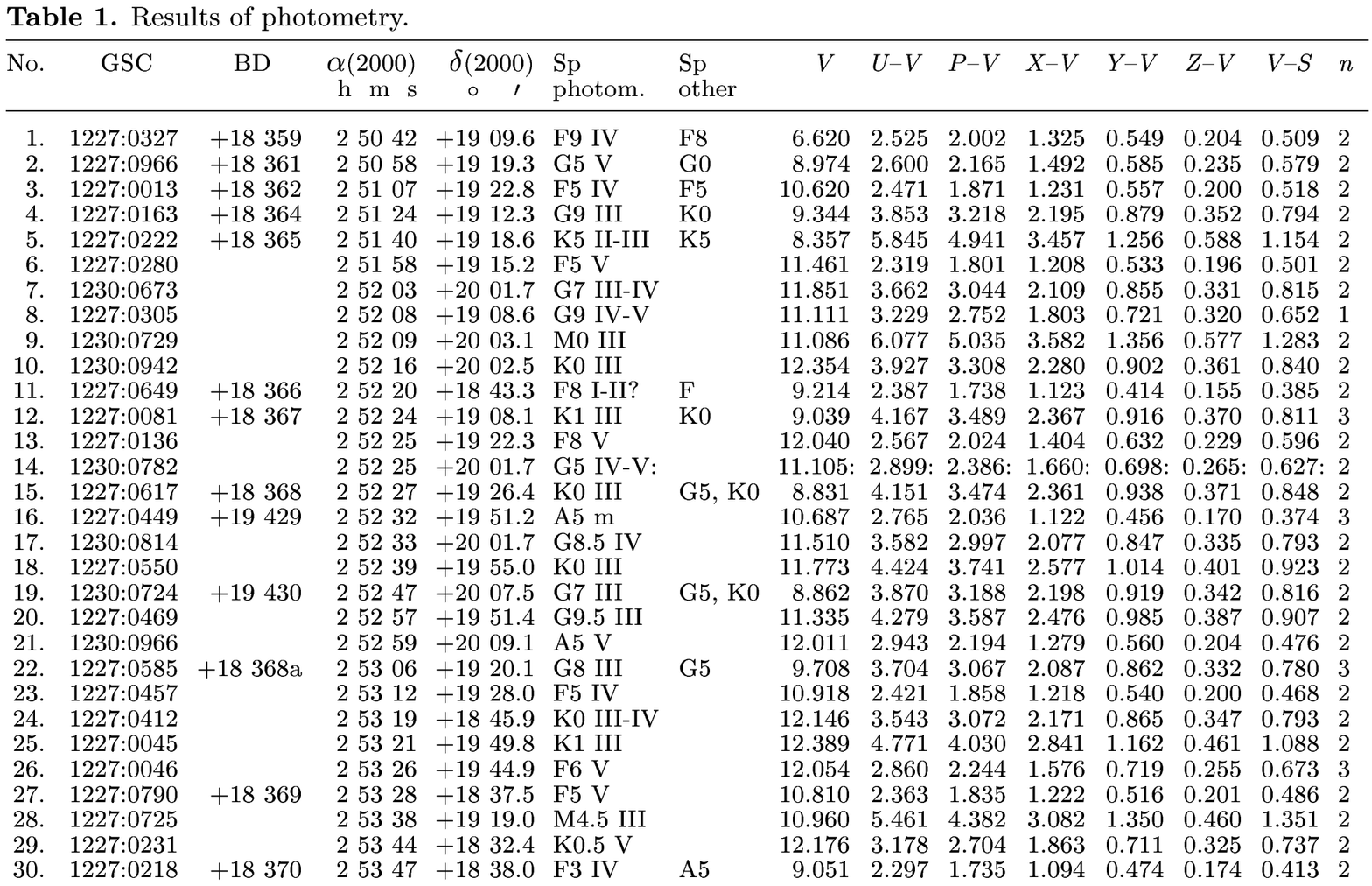,width=12.4truecm,angle=90,clip=}

\psfig{figure=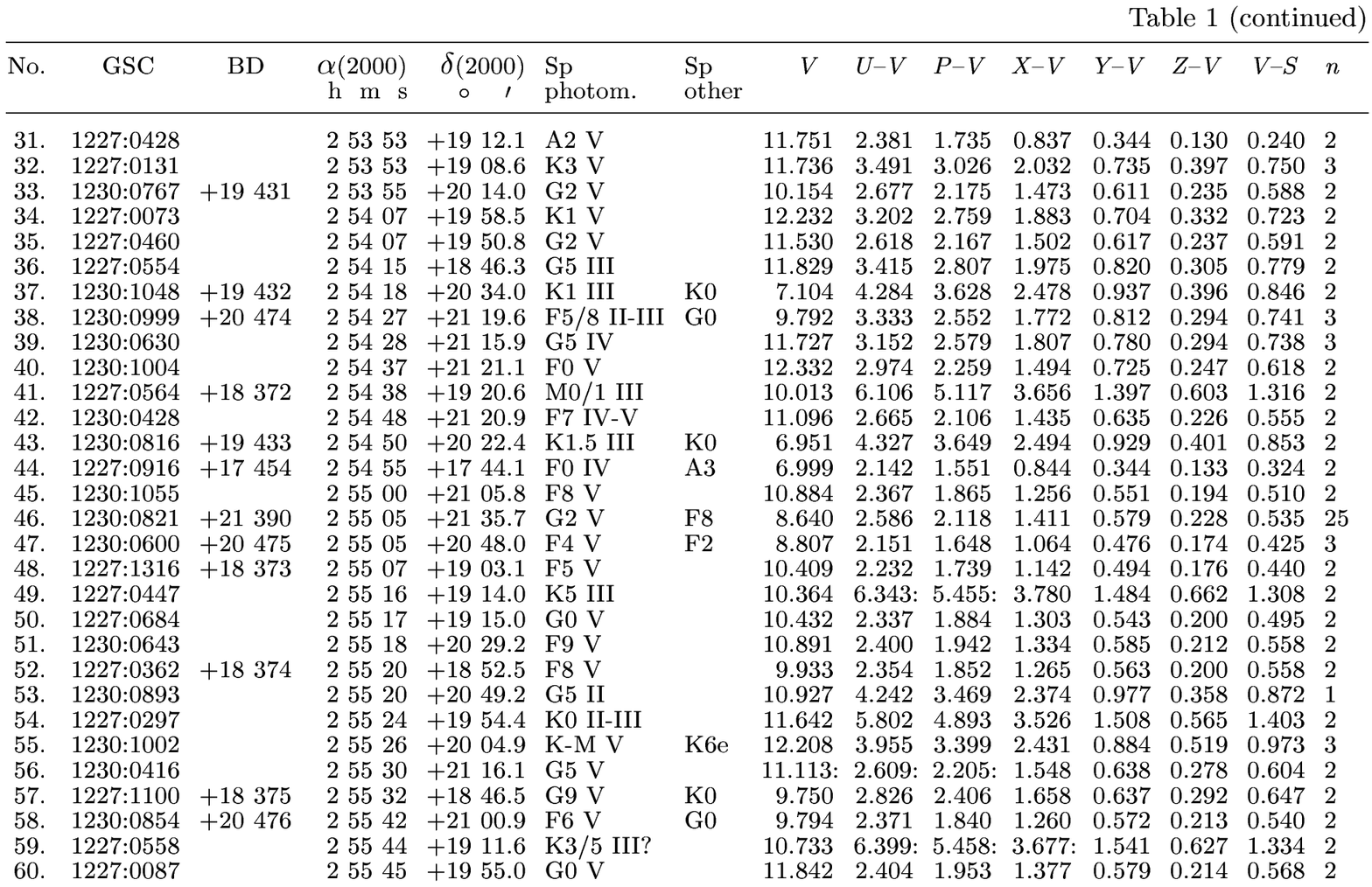,width=12.4truecm,angle=90,clip=}

\psfig{figure=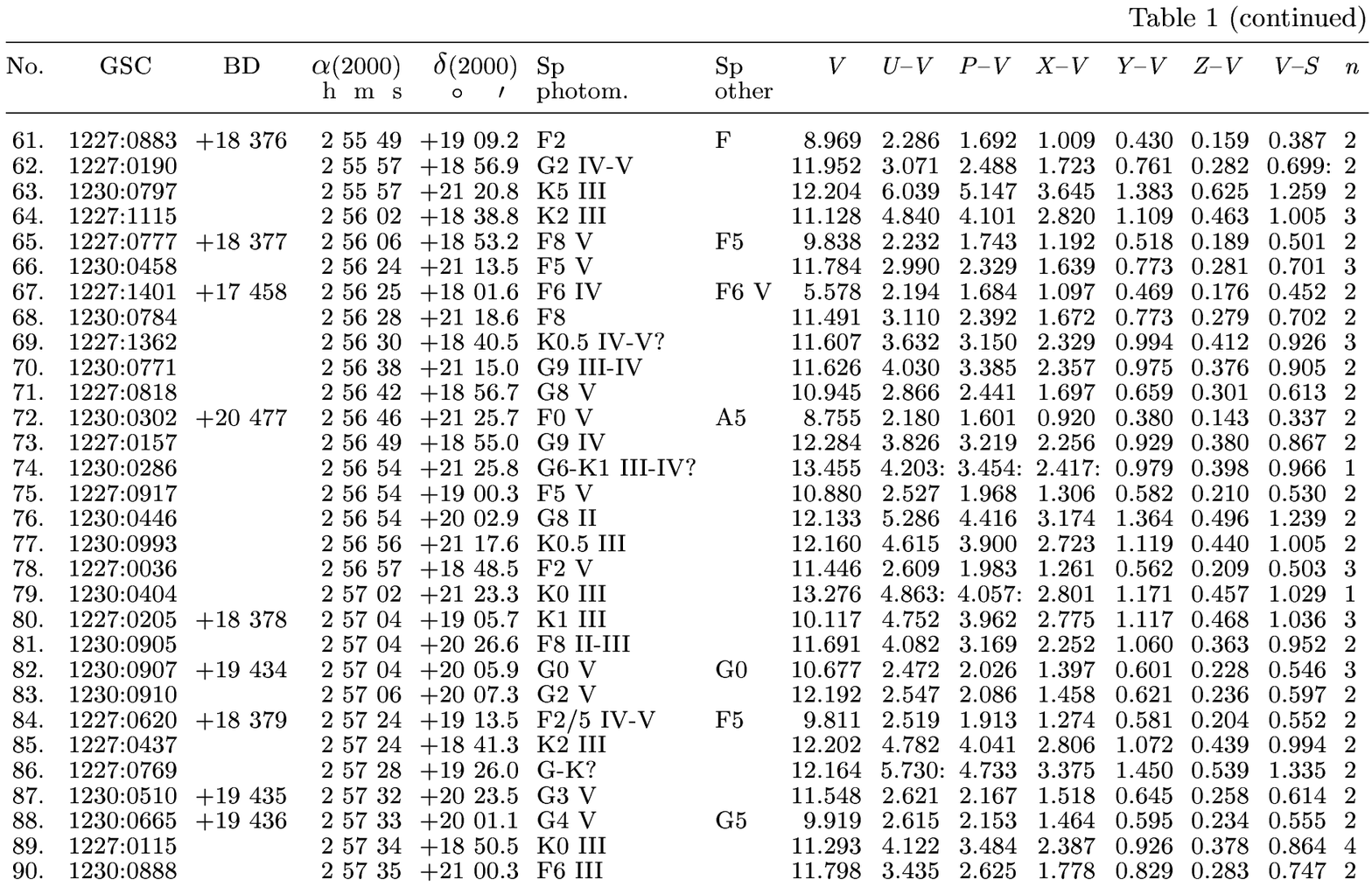,width=12.4truecm,angle=90,clip=}

\psfig{figure=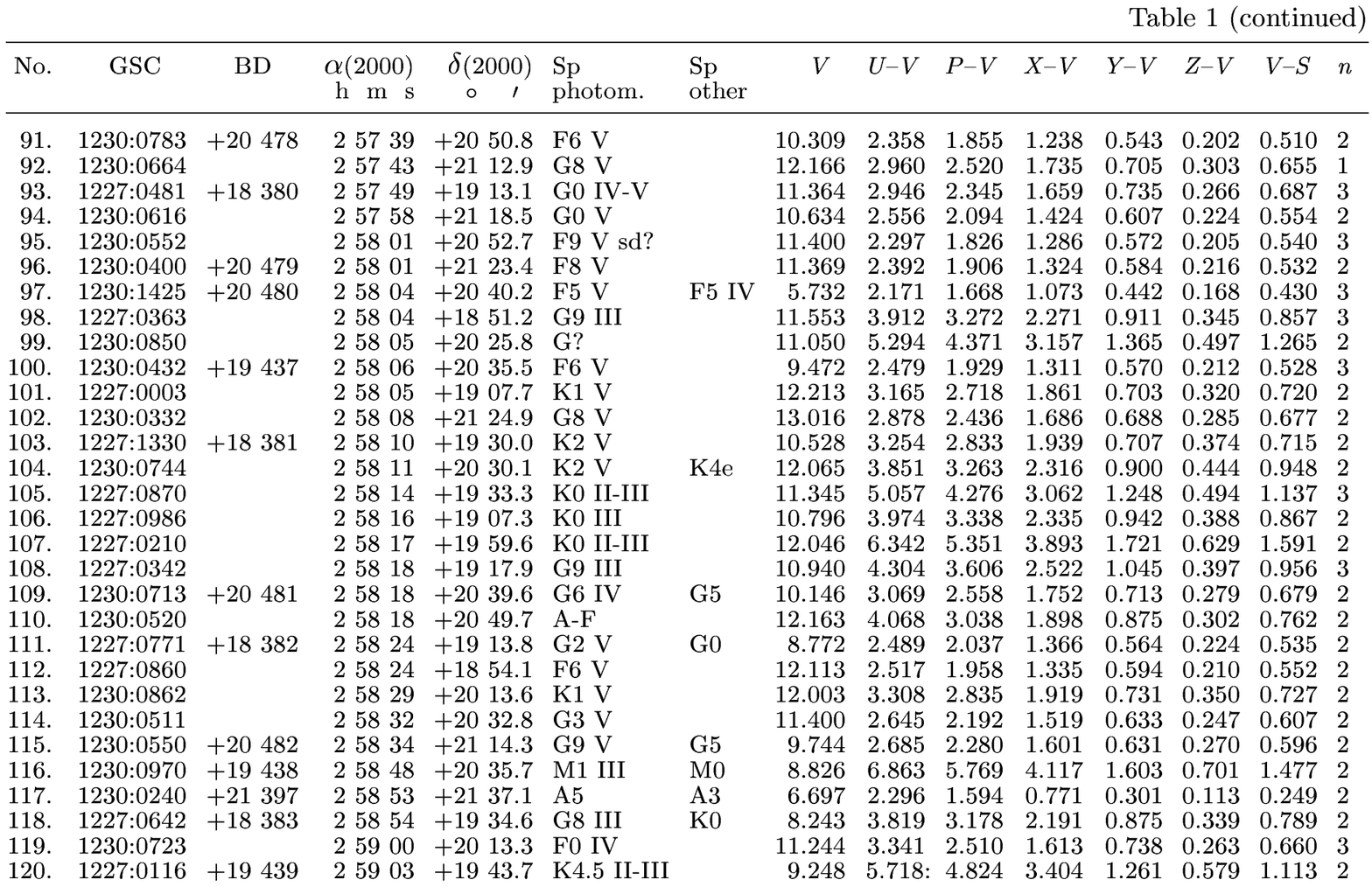,width=12.4truecm,angle=90,clip=}

\psfig{figure=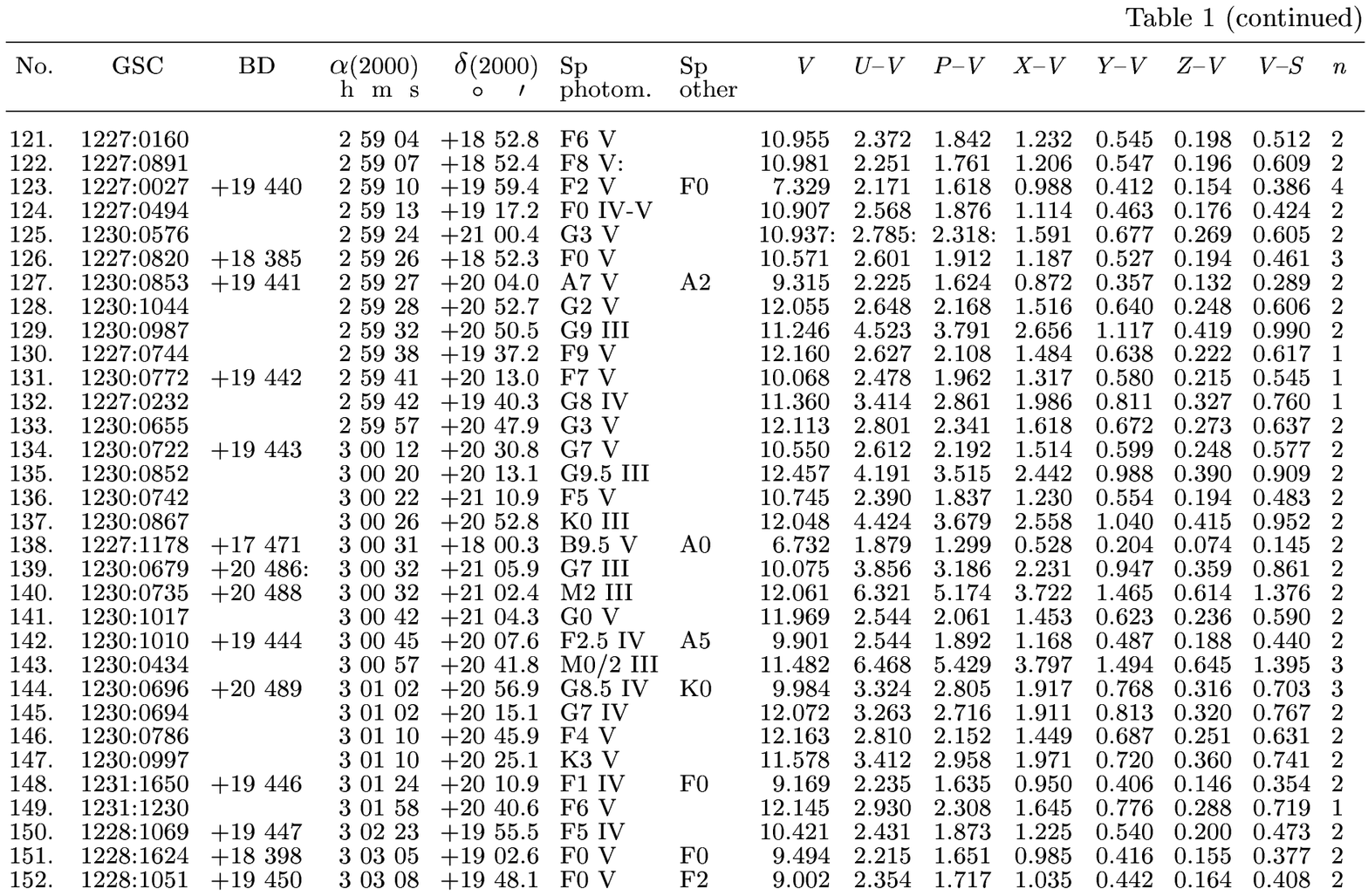,width=12.4truecm,angle=90,clip=}

\vfil\eject

\noindent NOTES:        \hfill\break
\vskip-3mm

{\tabfont

\noindent\hskip2pt~~1. HD 17659 (F8), HIC 13269. \hfil

\noindent\hskip2pt~~5. HD 17768 (K5), HIC 13337. \hfil

\noindent\hskip1.5pt~11. HD 17834 (F). \hfil

\noindent\hskip1.5pt~19. HD 17870 (G5), HIC 13418. \hfil

\noindent\hskip1.5pt~30. HIC 13496. \hfil

\noindent\hskip1.5pt~37. HD 18019 (K0), HIC 13532. \hfil

\noindent\hskip1.5pt~43. HD 18066 (K0), HIC 13571. \hfil

\noindent\hskip1.5pt~44. HD 18091 (A3), HIC 13579. \hfil

\noindent\hskip1.5pt~46. HD 18106 (F8), HIC 13589, standard star. \hfil

\noindent\hskip1.5pt~47. HD 18090 (F2). \hfil

\noindent\hskip1.5pt~57. HIC 13631. \hfil

\noindent\hskip1.5pt~61. HD 18190 (F), binary (see Paper II). \hfil

\noindent\hskip1.5pt~67. HD 18256 (F5), HIC 13702, $\rho$ Ari. \hfil

\noindent\hskip1.5pt~72. HD 18283 (A5), HIC 13723. \hfil

\noindent\hskip1.5pt~96. Binary (see Paper II). \hfil

\noindent\hskip1.5pt~97. HD 18404 (F0), HIC 13834. \hfil

\noindent 111. HIC 13855. \hfil

\noindent 117. HD 18484 (A3), HIC 13892, binary (see Paper II).\hfil

\noindent 118. HD 18485 (K0), HIC 13893. \hfil

\noindent 123. HD 18508 (F0), HIC 13913. \hfil

\noindent 131. Binary (see Paper II). \hfil

\noindent 138. HD 18654 (A0), HIC 14021, 50 Ari, binary (see Paper II).
\hfil

\noindent 152. HIC 14201. \hfil }

\bye

%% file: balticw.tex
%
%
\magnification=\magstep1
\baselineskip=11pt plus .1pt minus .1pt
\hsize=12.5truecm
\vsize=19.0truecm  
\hfuzz=5pt\vfuzz=5pt
\tolerance=1000
\overfullrule=0pt
\parskip=0pt
\abovedisplayskip=3 mm plus6pt minus 4pt
\belowdisplayskip=3 mm plus6pt minus 4pt
\abovedisplayshortskip=0mm plus6pt minus 2pt
\belowdisplayshortskip=2 mm plus4pt minus 4pt
\predisplaypenalty=0
\clubpenalty=10000
\widowpenalty=10000
\parindent=2em
%
%
\font\pgnumfont=cmr9
\font\headlinefont=cmti9
 \font\titlefont=cmbx10
\font\authorfont=cmr10
\font\addressfont=cmti9
\font\datefont=cmr9
\font\sumfont=cmr9
\font\itl=cmti9

\font\absfont=cmbx9
\font\secfont=cmr10
\font\subsecfont=cmti10
\font\subsubsecfont=cmr10
\font\figfont=cmr9
\font\figheadfont=cmbx9
\font\tabfont=cmr9
\font\tabheadfont=cmbx9
\font\mainfont=cmr10
\font\petitrm=cmr9

%
%
%
\newtoks\TITLE \newtoks\AUTHOR \newtoks\ADDRESS \newtoks\SUMMARY
\newdimen\sumindent \sumindent=\parindent
\newtoks\KEYWORDS \newtoks\SUBMITTED \newtoks\ACCEPTED
\newtoks\SENDOFF
%

%
%
\newtoks\firstpage
\let\firstpage=Y
\newtoks\AUTHORHEAD \newtoks\ARTHEAD \newtoks\VOLUME \newtoks\PAGES
\if!\the\AUTHORHEAD!\AUTHORHEAD={\the\AUTHOR}\fi
\if!\the\ARTHEAD!\ARTHEAD={\the\TITLE}\fi
\footline={\hfil}
\headline={\ifodd\pageno\rightheadline \else\leftheadline\fi}
\def\leftheadline{\if Y\firstpage\firsthead\global\let\firstpage=N
  \else\lefthead\fi}
\def\rightheadline{\if Y\firstpage\firsthead\global\let\firstpage=N
  \else\righthead\fi}
\def\lefthead{\pgnumfont\number\pageno\hfil\headlinefont\the\AUTHORHEAD}
\def\righthead{\headlinefont\the\ARTHEAD\hfil\pgnumfont\number\pageno}
\def\firsthead{\headlinefont Baltic Astronomy,~vol.\the\VOLUME,
\the\PAGES,~\the\year .\hfil}
\voffset=2\baselineskip 
%

\newdimen\oldbaselineskip \oldbaselineskip=\baselineskip
\def\test#1{\newlinechar=`@\if!\the#1! \message{#1 not given@}\fi}%
\def\printheader{
  \parindent=0pt
  \null\vskip1.cm
  \test{\TITLE}
  \vbox{\baselineskip=15pt
    \titlefont\the\TITLE
    }
  \vskip8mm plus8mm
  \test{\AUTHOR}
  \authorfont\the\AUTHOR
  \vskip2mm
  \test{\ADDRESS}
  \addressfont\the\ADDRESS
  \vskip2mm
  \test{\SUBMITTED}
  \line{\datefont Received \the\SUBMITTED
    \if!\the\ACCEPTED!\else, accepted \the\ACCEPTED\fi.\hfill}
  \vskip4mm plus4mm
  \vbox{\leftskip=\sumindent\parindent=0pt
    \parskip=5pt
    \absfont Abstract.
    \test{\SUMMARY}
    \sumfont\the\SUMMARY\par
    \absfont Key words:
    \test{\KEYWORDS}
    \sumfont\the\KEYWORDS\par
    }
  \sumfont
  \if!\the\SENDOFF!\else\footnote{}{Send offprint requests to:
 \the\SENDOFF}\fi
  \parindent=2em
  }
%
%
\newdimen\uppergap \newdimen\lowergap
\uppergap=5mm \lowergap=3mm
\newdimen\secind \newdimen\subsecind \newdimen\subsubsecind
\setbox0=\hbox{\secfont 9. }\secind=\wd0
\setbox0=\hbox{\subsecfont 9.9. }\subsecind=\wd0
\setbox0=\hbox{\subsubsecfont 9.9.9. }\subsubsecind=\wd0
\def\section#1{\goodbreak\par\vskip\uppergap
  \noindent\hangindent\secind\hangafter=1\secfont#1
  \vskip\lowergap\mainfont\par\nobreak}
\def\subsection#1{\goodbreak\par\vskip\uppergap
  \noindent\hangindent\subsecind\hangafter=1\subsecfont#1
  \vskip\lowergap\mainfont\par\nobreak}
\def\subsubsection#1{\goodbreak\par\vskip\uppergap
  \noindent\hangindent\subsubsecind\hangafter=1\subsubsecfont#1
  \vskip\lowergap\mainfont\par\nobreak}
%
%
\def\WFigure#1#2#3{\goodbreak\midinsert\vbox{
  \null\centerline{#2}\vskip5truemm
  \figheadfont\indent Fig.~#1.\figfont\ #3
  \par\mainfont
  }\endinsert}
%

%

%

%

%
\newdimen\tabind
\setbox0=\hbox{\tabheadfont Table 55.} \tabind=\wd0

%
%
\def\References{\vskip\uppergap
\line{\secfont REFERENCES\hfill}
  \vskip0.8\lowergap
 \petitrm
  }
\def\ref{\goodbreak
\hangindent12pt\hangafter=1
\noindent\ignorespaces}
\def\endref{\egroup}
%
%
\def\byebye{\egroup\par\vfill\supereject\end}
%
%

%
%

\def\degr{\hbox{$^\circ$}}

\def\arcmin{\hbox{$^\prime$}}

\def\utw{\smash{\rlap{\lower5pt\hbox{$\sim$}}}}
\def\udtw{\smash{\rlap{\lower6pt\hbox{$\approx$}}}}


\font\tabfont=cmr9



\def\ddown{\lower2.5ex\hbox}
\def\ddow{\lower1.7ex\hbox}
\def\down{\lower1ex\hbox}
\def\uppp{\raise1ex\hbox}
\def\dnnn{\lower1ex\hbox}
\def\uuppp{\raise2ex\hbox}

\def\ts{\thinspace}
\def\(o-c){$O-C$}


\def\angstr{A\kern-.56em\raise1.9ex\hbox{$\scriptscriptstyle\circ$}$\,$}

\newdimen\free\newdimen\shift
\def\Entry#1#2#3{\par\goodbreak\smallskip%
  \setbox1=\vbox{\advance\hsize by-10mm\parindent=0pt
    \def\\{\par}%
    \it#1. \rm#2}
  \line{\box1\hfill#3}\smallskip
}%
\newdimen\savesize

\def\shiftfigure #1#2#3#4#5{
    \vbox to #2 { \ifodd #5 \rightskip#4 \else\leftskip#4 \fi
                  \null\vfil
                  \figheadfont Fig.~#1.\figfont #3
                  \medskip
                }
                          }

\year2002